\documentclass[twocolumn,showpacs,preprintnumbers,amsmath,aps,prl,longbibliography,amssymb,superscriptaddress]{revtex4-1}

\usepackage[english]{babel}
\usepackage{braket}
\usepackage{amsmath}
\usepackage{amssymb}
\usepackage{commath}
\usepackage{siunitx}
\usepackage[latin1]{inputenc}
\usepackage{graphicx}
\usepackage{tikz}
\usepackage[caption=false]{subfig}
\usepackage[colorlinks,bookmarks=true,citecolor=blue,linkcolor=red,urlcolor=blue]{hyperref}

\let\vec\mathbf
\newcommand{\state}[1]{\ensuremath{{| #1 \rangle}}}

\begin{document}

\title{Interaction-driven plateau transition between integer and fractional Chern Insulators}

\author{Leon Schoonderwoerd} 
\affiliation{Physics and Astronomy, Division of Natural Sciences, University of Kent, Canterbury CT2 7NZ, United Kingdom}

\author{Frank Pollmann}
\affiliation{Department of Physics, TFK, Technische Universit\"at M\"unchen, James-Franck-Stra{\ss}e 1, D-85748 Garching, Germany}
\affiliation{Munich Center for Quantum Science and Technology (MCQST), Schellingstra{\ss}e 4, D-80799 M\"unchen, Germany}

\author{Gunnar M\"oller}
\affiliation{Physics and Astronomy, Division of Natural Sciences, University of Kent, Canterbury CT2 7NZ, United Kingdom}

\begin{abstract}
We present numerical evidence of an interaction-driven quantum Hall plateau transition between a $|C|>1$ Chern Insulator (CI) and a $\nu = 1/3$ Laughlin state in the Harper-Hofstadter model. 
We study the model at flux densities $p/q$, where the lowest Landau level (LLL) manifold comprises $p$ magnetic sub-bands. 
For weak interactions, the model realises integer CIs corresponding to filled sub-bands, while strongly interacting candidate states include fractional quantum Hall (FQH) states at LLL filling fractions $\nu=r/t$. 
These phases may compete at the same particle density when $p=t$.
As a concrete example, we numerically explore the physics at flux density $n_{\phi} = 3/11$, where we show evidence that a direct transition occurs between a CI and a $\nu = 1/3$ Laughlin state, which we characterise in terms of its critical, topological and entanglement properties.
We also show that strong interactions generically stabilise a $\nu = 1/3$ Laughlin state even when the LLL is split into multiple bands, and introduce a powerful methodology to extract its topological entanglement entropy by exploiting the scaling of magnetic length with $n_\phi$. 
\end{abstract}

\maketitle

Lattice fractional quantum Hall (FQH) effects have received intense recent interest, owing to proposals and subsequent successful generation of topological band structures in cold atomic gases \cite{linSyntheticMagneticFields2009, aidelsburgerRealizationHofstadterHamiltonian2013, miyakeRealizingHarperHamiltonian2013, aidelsburgerMeasuringChernNumber2014a, jotzuExperimentalRealizationTopological2014, cooperRapidlyRotatingAtomic2008, dalibardColloquiumArtificialGauge2011, goldmanLightinducedGaugeFields2014}, and using superlattice structures in graphene systems \cite{deanHofstadterButterflyFractal2013, spantonObservationFractionalChern2018, hensgensCapacitanceSpectroscopybasedPlatform2018, forsytheBandStructureEngineering2018, knappFractionalChernInsulator2019}.
Similar prospects exist in 2D materials with strong spin-orbit coupling \cite{tangHighTemperatureFractionalQuantum2011, neupertFractionalQuantumHall2011, sunNearlyFlatbandsNontrivial2011}, although no concrete materials have yet been identified. 
All of these realizations can provide topological flat bands with an effective magnetic length comparable to the lattice spacing. 
When combined with strong repulsive interactions, such systems may exhibit fractional Chern insulator (FCI) states: FQH liquids stabilised specifically by a lattice potential \cite{kolFractionalQuantumHall1993, Moller2009, mollerFractionalChernInsulators2015, spantonObservationFractionalChern2018}. 
While literature on FCIs focuses on models where a single low-lying flat topological band mirrors the physics of the lowest Landau level (LLL) \cite{tangHighTemperatureFractionalQuantum2011, neupertFractionalQuantumHall2011, sunNearlyFlatbandsNontrivial2011, Regnault2011, bergholtzTopologicalFlatBand2013, parameswaranFractionalQuantumHall2013a}, composite fermion (CF) theory approaches lattice quantum Hall (QH) effects from a more general angle, and predicts incompressible states even when the LLL is split into numerous sub-bands \cite{jainCompositefermionApproachFractional1989, lopezFractionalQuantumHall1991, kolFractionalQuantumHall1993, Moller2009}.

QH states are inherently interesting for fundamental physics as they can support exotic fractionalized quasiparticles obeying non-Abelian exchange statistics \cite{Moore1991} and could be exploited as a platform for universal topological quantum computation \cite{nayakNonAbelianAnyonsTopological2008}. 
Transitions between QH states have likewise attracted much attention \cite{weiExperimentsDelocalizationUniversity1988, engelCriticalExponentFractional1990, huckesteinScalingTheoryInteger1995, sondhiContinuousQuantumPhase1997}. 
Conventionally, QH plateau transitions are driven by field strength tuning the Landau level filling.
Their critical properties can be understood in terms of the percolation of current-carrying modes \cite{huckesteinScalingTheoryInteger1995}, captured by network models \cite{chalkerPercolationQuantumTunnelling1988} or renormalisation group approaches \cite{khmelnitskiiQuantizationHallConductivity1983, kivelsonGlobalPhaseDiagram1992}. 
Several scenarios for transitions into or between FQH states have also been described in clean systems \cite{wenTransitionsQuantumHall1993,chenMottTransitionAnyon1993,yeCoulombInteractionsQuantum1998, groverQuantumPhaseTransition2013, barkeshliContinuousTransitionFractional2014, barkeshliContinuousPreparationFractional2015, leeEmergentMultiFlavorQED2018}. 
These are often cast as Chern number changing transitions in the underlying CF \cite{leeEmergentMultiFlavorQED2018} or parton description \cite{barkeshliContinuousTransitionFractional2014}. 
The critical point is then described by a  massless multi-flavour Dirac theory coupled to a gauge field \cite{groverQuantumPhaseTransition2013, barkeshliContinuousTransitionFractional2014, luQuantumPhaseTransitions2014, leeEmergentMultiFlavorQED2018, maEmergentMathrmQCDQuantum2020}. 
A microscopic realization was found in graphene heterostructures, where transitions between FCI states within the same Jain series are driven by tuning the strength of a periodic potential \cite{leeEmergentMultiFlavorQED2018}.

This manuscript explores interaction-driven QH plateau transitions in the Harper-Hofstadter model \cite{harperSingleBandMotion1955a, azbelEnergySpectrumConduction1964a, hofstadterEnergyLevelsWave1976a}, motivated partly by its recent experimental realization  \cite{deanHofstadterButterflyFractal2013, aidelsburgerRealizationHofstadterHamiltonian2013, miyakeRealizingHarperHamiltonian2013, aidelsburgerMeasuringChernNumber2014a}. 
We study flux densities $n_{\phi}$ where the LLL is split into several magnetic sub-bands to allow competition between integer sub-band filling and fractional filling of the LLL. 
We show that repulsive interaction strengths can drive transitions between integer and fractional CIs in the Harper-Hofstadter model. 
Specifically, we provide numerical evidence suggesting a direct transition between a $C=+4$ Chern insulator (CI) and a $\nu=1/3$ Laughlin state \cite{laughlinAnomalousQuantumHall1983a}, at $n_{\phi}=3/11$, where the LLL comprises three sub-bands. 
These two states are members of different Jain series, as the flux attached to CFs changes. 
Hence, this interaction-driven plateau transition cannot be described as CF Chern number changing, and the previously considered critical theories \cite{barkeshliContinuousTransitionFractional2014, leeEmergentMultiFlavorQED2018} do not apply.

\begin{figure}
\begin{tikzpicture}
    \node at (0.1\linewidth, 0) {\resizebox{0.38\linewidth}{!}{\includegraphics{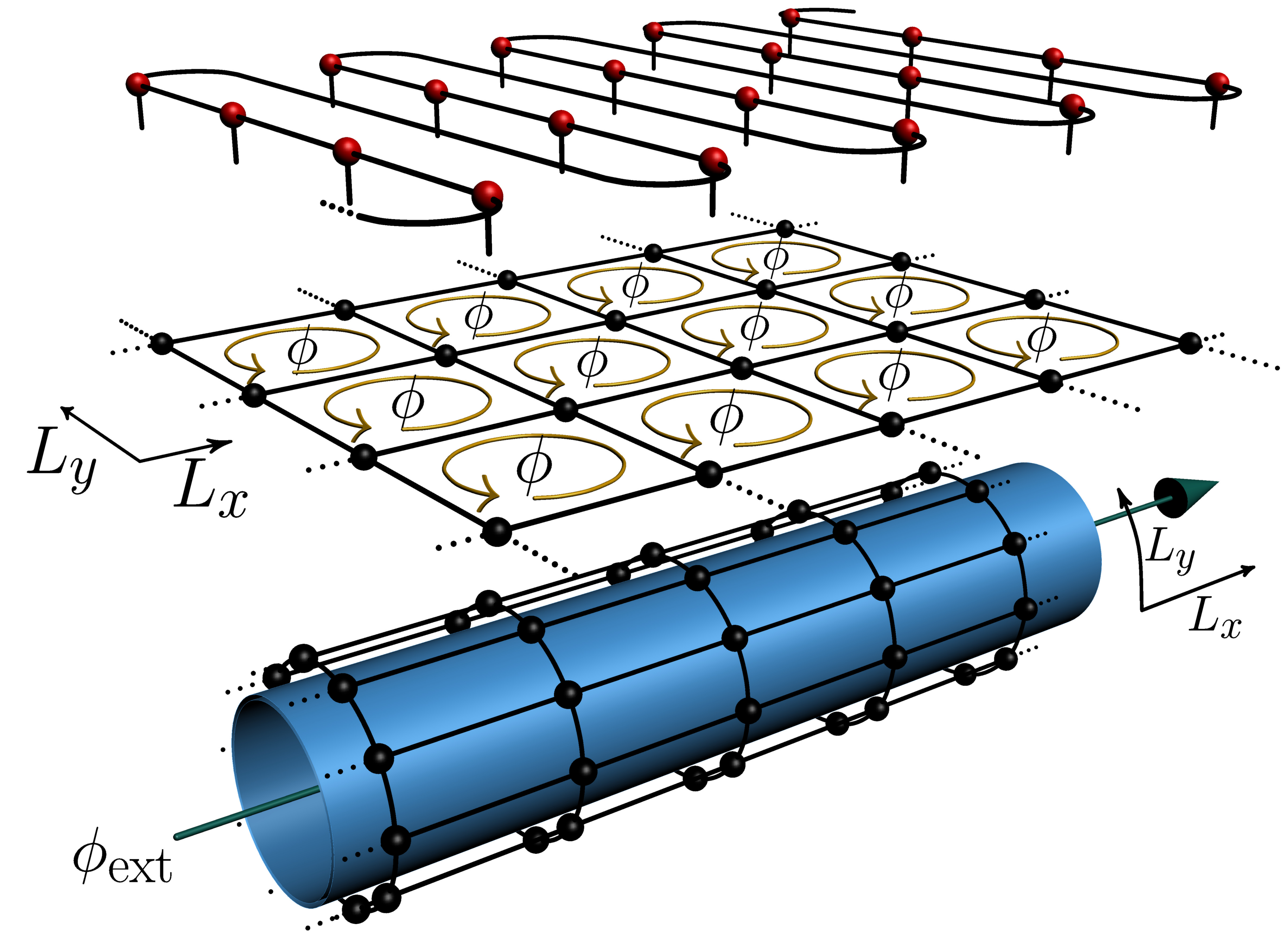}}}; 
    \node at (0.58\linewidth, 0) {\resizebox{0.45\linewidth}{!}{\includegraphics{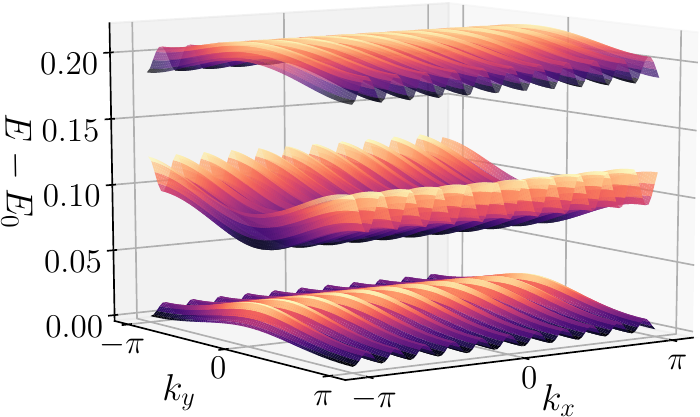}}}; 
    \node at (-0.05\linewidth,0.16\linewidth) [scale=0.8,label={[name=label node]below left:$(a)$}] {}; 
    \node at (0.4\linewidth,0.16\linewidth) [scale=0.8,label={[name=label node]below left:$(b)$}] {};
  \end{tikzpicture}
	\caption{(a) The Harper-Hofstadter model in the Landau gauge with constant flux per plaquette (center) is `wrapped' around the cylinder (bottom), 
	and corresponding Matrix Product State representation (top).
	(b) Single-particle band structure at $n_{\phi} = 3/11$ (lowest $3$ bands).
	\label{fig:hofstadter_combined} }
\end{figure}

Most studies of lattice FQH effects or FCIs have studied regimes where the low-energy manifold of states consists of a single band \cite{royBandGeometryFractional2014, mollerFractionalChernInsulators2015, bauerQuantumGeometryStability2016, andrewsStabilityFractionalChern2018, Regnault2011, scaffidiAdiabaticContinuationFractional2012a, jacksonGeometricStabilityTopological2015, bergholtzTopologicalFlatBand2013, parameswaranFractionalQuantumHall2013a, Andrews:2021it, Andrews:2021bb}, simplifying the numerical treatment. 
Here, we explicitly consider situations where the LLL is split into multiple sub-bands. 
This situation arises naturally in the Harper-Hofstadter model (Fig.~\ref{fig:hofstadter_combined}), owing to its self-similar bandstructure \cite{hofstadterEnergyLevelsWave1976a, splittingChernBands}. 
The model is described by the fermionic tight-binding Hamiltonian
\begin{equation} \label{eq:hofstadter_hamiltonian}
\mathcal{\hat{H}}_{\mathrm{Hof}} = - t \sum_{\langle ij \rangle} \left[ e^{i \phi_{ij}}  \hat{a}^{\dagger}_{i} \hat{a}_{j} + \mathrm{H.c.}\right] + V \sum_{\langle ij \rangle} \hat{n}_{i} \hat{n}_{j},
\end{equation}
with hopping amplitude $t$, nearest-neighbour repulsion $V$, annihilation (creation) operators $\hat{a}$ ($\hat{a}^{\dagger}$), the number operator $\hat{n}=\hat{a}^{\dagger}\hat{a}$ and $\phi_{ij} = \int^{i}_{j} \vec{A} \cdot \dif \vec{l}$ are Aharonov-Bohm phases on $\langle i, j \rangle$. We choose the Landau gauge with vector potential $\vec{A} = xB\hat{\vec{y}}$ to yield a homogeneous magnetic field strength $B$ with flux density $n_{\phi}=B/\Phi_0$ in units of the magnetic flux quantum $\Phi_0 = \frac{h}{e}$.

At rational $n_\phi=p/q$, the single-particle Hofstadter spectrum consists of $q$ bands, each with areal density of states $n_\text{band}=1/q$. 
The largest gaps arise at particle density $n=n_\phi$, and are connected continuously to LLL physics in the limit of $n_\phi \rightarrow 0$ \cite{hofstadterEnergyLevelsWave1976a, wannierResultNotDependent1978}. 
As a corollary, for general $n_\phi$ this LLL is split into $p$ bands. 
All other gaps in the spectrum can be considered as fractal copies of the principal gap in subordinate instances of the self-similar spectrum \cite{hofstadterEnergyLevelsWave1976a}. 
Moreover, all bands are topological, carrying a non-zero Chern number which can be inferred from the diophantine relations, $n = C n_\phi + s$ \cite{stredaTheoryQuantisedHall1982, thoulessQuantizedHallConductance1982a}.

Non-interacting fermions filling the lowest $m$ bands in the spectrum yield a CI state with integer Hall conductivity $\sigma_H = e^2/h \sum_{i=1}^m C_i$ \cite{stredaTheoryQuantisedHall1982, thoulessQuantizedHallConductance1982a, niuQuantizedHallConductance1985a, osadchyHofstadterButterflyQuantum2001}. 
However, with increasing $V>0$, interaction energy starts to dominate over kinetic energy, and can no longer be treated perturbatively on the non-interacting ground-state. 
Consequently, the system favours larger inter-particle separation at the expense of lifting particles beyond the lowest-lying bands. 
The question now is which magnetic bands contribute significantly to the ground state interacting problem. 
CF theory of the Harper-Hofstadter model posits a regime dominated by a low-energy manifold of $m_s = q n_s $ bands located below a sufficiently large single-particle gap \cite{kolFractionalQuantumHall1993, Moller2009,  mollerFractionalChernInsulators2015}, enabling the prediction of candidate FCI states with filling fraction in the low-energy manifold $\nu = n/n_s=r/(kCr + 1)$, with $k,r \in \mathbb{Z}$ \cite{mollerFractionalChernInsulators2015}. 
For flux densities $n_\phi$ allowing $m_s = |kCr + 1|$ bands in the low-energy manifold, the CI filling the lowest $m$ bands can compete directly with an FCI realised at the same particle density, but involving states in the lowest $m_s$ bands.
In principle, larger interaction strength could lead to further mixing of higher-lying bands, destabilising the CF FQH state.

Concretely, we consider the simplest example of such a transition between a CI and a $\nu=1/3$ Laughlin state in the $C=1$ LLL manifold ($k=2$, $r=1$). 
We require $m_s = |2 \times 1 \times 1 + 1|  = 3$ sub-bands in the LLL, thus $n_\phi=3/q$, or $p=m_s=3$. 
We further specialise to an example  with $n_\phi=3/11$, where the lowest sub-band has a Chern number $C_1=4$.

We numerically analyse the competition between the candidate states using infinite Density Matrix Renormalization Group (iDMRG) methods \cite{whiteDensityMatrixFormulation1992, McCulloch2008},
expressing (\ref{eq:hofstadter_hamiltonian}) in a Matrix Product Operator (MPO) representation [Fig.~\ref{fig:hofstadter_combined}(b)] \cite{schollwockDensitymatrixRenormalizationGroup2011c, hauschildEfficientNumericalSimulations2018} for the well-established infinite-cylinder lattice geometry \cite{Yan2011, Motruk2015, motrukDensityMatrixRenormalization2016a, gersterFractionalQuantumHall2017, motrukPhaseTransitionsAdiabatic2017a}. 
Entanglement grows linearly with cylinder circumference, limiting the maximum circumference $L_y$ that can be accurately approximated at a given Matrix Product State (MPS) bond dimension $\chi$.

\begin{figure}[tp]
	\includegraphics[width=\columnwidth]{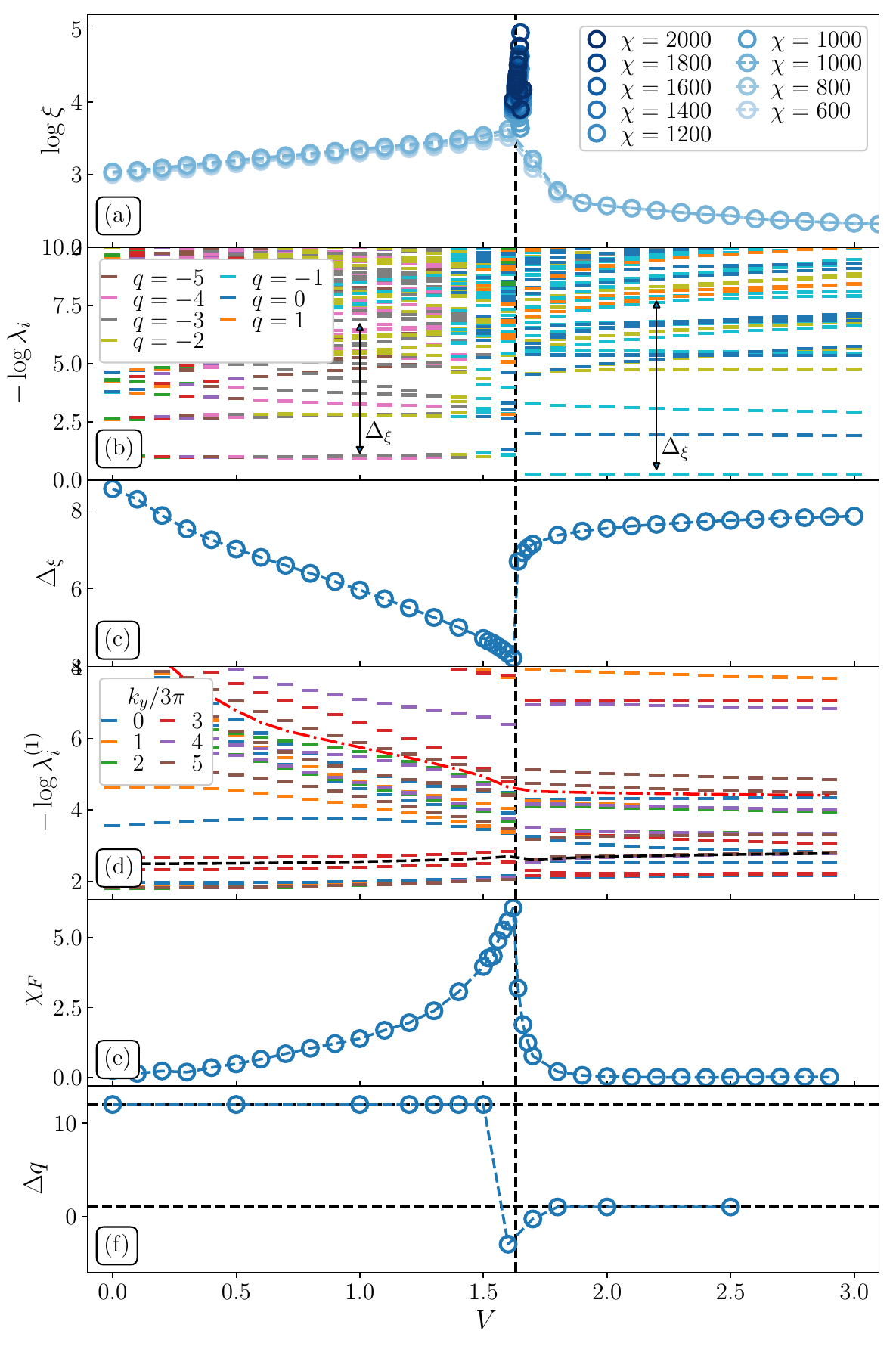} 
	\caption{Several observables through the transition (dashed line).  
		All results are for $L_y = 6$, $n_{\phi} = 3/11$, $\chi=500$ [except $600 \leq \chi \leq 1000$ in (a)]. 
		(a)~Correlation length for several values of the bond dimension. 
		(b)~Momentum-space entanglement spectrum (colors denote different quantum number sectors). Also shows the entanglement gap.
		(c)~Entanglement gap. 
		(d)~Spectrum of the single-particle density matrix. The dashed line indicates the lowest $6$ ($18$) states on the left (right) of the transition.
		(e)~Fidelity susceptibility $\chi_F =  (|\langle \psi (V) | \psi (V+ \delta V) \rangle |^2 - 1) / (\delta V)^2$.
		(f)~Charge $\langle \Delta q \rangle$ transported along the cylinder by the insertion of $3 \times 2 \pi$ flux.
		\label{fig:transition_observables}}
\end{figure}

Fig.~\ref{fig:transition_observables} summarizes the evolution of several observables as a function of interaction strength $V$, for our target scenario: $n_\phi=3/11$, $L_y = 6$. 
For additional data for $L_y = 7$ and $L_y = 8$, see Appendix D. 
We find multiple indications of a direct CI-FCI phase transition at a critical $V_c$.

The first sign of a phase transition is found in the correlation length $\xi$.
At $V_c$, our data for $\xi(\chi)$ suggest a divergence with bond dimension [Fig.~\ref{fig:transition_observables}(a)].
This is consistent with a continuous quantum phase transition occurring at $V_c$. 
The entanglement spectrum (ES) $\lbrace - \log \lambda_i(k_y) \rbrace$ of the Schmidt decomposition into two half-infinite states $\state{\Psi}=\sum_{i,k_y}\lambda_{i}(k_y) \state{\chi^L_{i,k_y}}\otimes \state{\chi^R_{i,k_y}}$ is shown in Fig.~\ref{fig:transition_observables}(b).
At $V_c$, it displays a sudden opening of the lowest gap, as well as other discontinuous changes in higher entanglement energies. 
In the ES, we define the entanglement gap $\Delta_{\xi}$ as the gap between the lowest two levels within the same quantum number and momentum sector. 
Fig.~\ref{fig:transition_observables}(c), shows a discontinuity of $\Delta_{\xi}$ at $V_c$, and a tendency of the gap to decrease for $V < V_c$ approaching the transition. 
Thus, the entanglement properties of the system also support a phase transition.

Fig.~\ref{fig:transition_observables}(d) displays the single-particle density matrix $\rho^{(1)}_{ij}=\langle \hat c^\dagger_j \hat c_i \rangle$ calculated within our MPS unit cell of $1\times 6$ magnetic unit cells, yielding six $k$-points in $k_y$-space.
In the CI phase, its spectrum is dominated by six low-lying eigenvalues and a seventh nearby (including at least one state at each $k$-point), while in the FCI phase a gap opens above $18$ low-lying states and two further nearby states (including at least three states at each $k$-point). 
This matches the expectation that only states in the lowest band contribute significantly to $\rho^{(1)}_{ij}$ in the CI phase, while it involves all states in the LLL (i.e., three bands) for the Laughlin state \cite{sterdyniakParticleEntanglementSpectra2012}.
As a function of $V$, $\rho^{(1)}_{ij}$ changes discontinuously between the CI and Laughlin phases at $V_c$, signalling a direct transition.
Fig.~\ref{fig:transition_observables}(e) shows the fidelity susceptibility $\chi_F =  (|\braket{ \psi (V) | \psi (V+ \delta V) } |^2 - 1) / (\delta V)^2$, which peaks at $V_c$, also supporting a direct phase transition.

Finally, we measure the Hall conductance using flux insertion \cite{laughlinQuantizedHallConductivity1981, thoulessQuantizedHallConductance1982a}.
We calculate the ground state's evolution under an adiabatic insertion of $3 \times 2 \pi$ external flux through the cylinder, and measure the charge $\Delta q$ thereby transported along the cylinder \cite{zaletelFluxInsertionEntanglement2014a,  grushinCharacterizationStabilityFermionic2015, motrukDensityMatrixRenormalization2016a, motrukPhaseTransitionsAdiabatic2017a},
which yields the Hall conductivity via $\sigma_{H} = \Delta q / \Delta \phi_{\mathrm{ext}}$.
Flux insertion for $V=0$ confirms that the system realizes an integer CI state with Hall conductivity proportional to the Chern number of the lowest Harper-Hofstadter band, $\sigma_H \propto C_1=4$.
Fig.~\ref{fig:transition_observables}(f) shows that this picture persists for small $V>0$.
At interaction strengths $V>V_c$, we find  $\sigma_H = 1/3$, consistent with the prediction for a $\nu = 1/3$ Laughlin state.
The Hall conductivity $\sigma_H(V)$ discontinuously jumps between these two extremal values. 
However, near $V_c$ we cannot reliably perform the flux pumping procedure. 
Here, the insertion violates adiabaticity, evidenced by discontinuities in $q(\phi)$. 
This is understood by the increase of the correlation length near $V_c$, requiring larger bond dimensions to fully capture the many-body state.

In sum, these multiple signatures clearly demonstrate a direct transition from the $\sigma_H=4$ CI state to a $\sigma_H=1/3$ Laughlin state. 
Near the transition, we see that our iDMRG calculation is not fully converged in the bond dimension $\chi$, highlighted by the continued increase in correlation length with $\chi$ in Fig.~\ref{fig:transition_observables}(a). 
This behaviour would be consistent with a continuous transition. 
Contrarily, the state's entanglement structure appears to evolve discontinuously at the transition, suggesting a (weakly) first order transition. 
As iDMRG on cylinders is incapable to unambiguously distinguish the nature of a quantum phase transition in two dimensions, our data could reflect either scenario. 
Further indications could be found by examining the evolution of the ground state degeneracy in a torus geometry.

While the transition at $L_y = 6$ is sharply defined even at low $\chi$, features at  higher-$L_y$ are more challenging to resolve using DMRG. At $L_y = 7$ a sharp feature in $\xi$ and $S$ emerges only between $2000\leq \chi \leq  3000$ while we only resolved a broader maximum for $L_y = 8$ up to $\chi=2000$. Additionally, the ES and spectrum of $\rho_{ij}$ vary more smoothly rather than discontinuously with $V$ when examined at low $\chi=600$, as shown in Appendix D. 
However, all three circumferences clearly support distinct phases at low and high $V$.

To identify possible theories describing the system at criticality, we extract the central charge for a putative conformal field theory from the simultaneous scaling of $S_E$ and $\xi$ with $\chi$.
These are expected to follow the relation $S_E(\chi) \sim (c/6) \log \xi(\chi)$ \cite{Calabrese2004}, where $c$ is the central charge, under finite entanglement scaling.
This procedure was originally developed for 1D systems \cite{tagliacozzoScalingEntanglementSupport2008, pollmannTheoryFiniteentanglementScaling2009}, but has subsequently been applied successfully to 2D systems \cite{geraedtsHalffilledLandauLevel2016, gohlkeDynamicsKitaevHeisenbergModel2017}.
We show numerical scaling data for $V \sim V_c$ and $L_y = 6, 7, 8$ in Fig.~\ref{fig:central_charge}(a), alongside the linear scaling predicted for $c = 1, 2$ and $3$ (additionally, see Appendix D for underlying data).

The excellent fits show a rapid increase of $c$ with $L_y$,
\begin{figure}[tp]
	\includegraphics[width=\columnwidth]{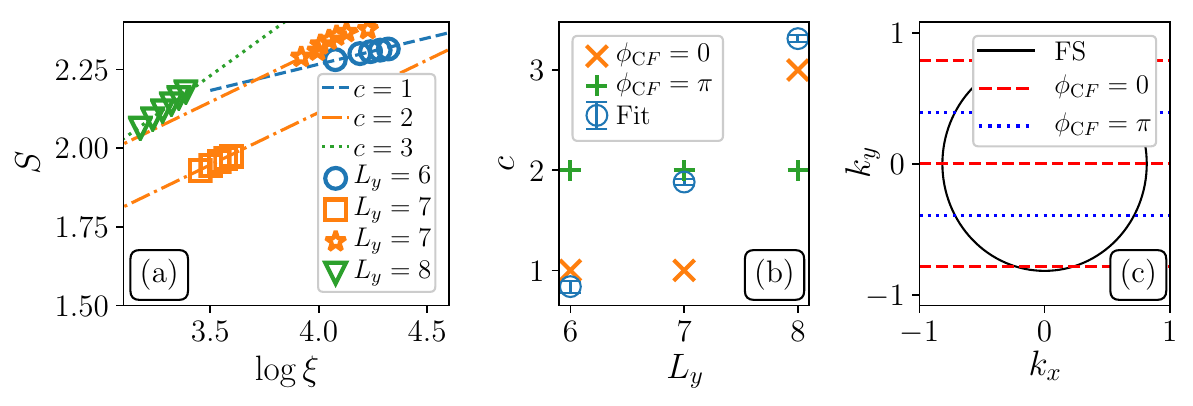} 
	\caption{
		(a) Scaling of $S_E$ and $\xi$ with $1000 \leq \chi \leq 2000\,(3000)$ for $L_y=6$: blue circles, $L_y =7$: orange squares (and stars) and $L_y = 8$: green triangles, at $V=1.634,$ $1.5,$ (and $1.55)$, and $1.45$ respectively, and expected linear scaling for $c=1$, $2$, $(2)$ and $3$, respectively.
		Dashed black lines denote linear fits.
		(b) $c (L_y)$, as inferred from linear fit to data (blue circles) and predicted for CF Fermi surface with $\phi_\text{CF}=0$ (orange crosses) and  $\phi_\text{CF}=\pi$ (green pluses).
		Error bars denote one standard deviation $\sigma$.
		(c) View of a putative Fermi surface for $L_y=8$, showing intersections with three (two) lines of allowed $k_y$ momentum for $\phi_\text{CF}=0 \,(\pi)$.
	}
	\label{fig:central_charge}
\end{figure}
which suggests that the critical system may be described by a Fermi Surface (FS) theory.
Following \cite{geraedtsHalffilledLandauLevel2016}, and assuming a circular FS with radius $k_F \ell_B = \sqrt{2/3}$, we count the expected numbers of `wires' $N_w$ cutting the FS (due to the quantisation of $k_y = \frac{1}{L_y} (2 \pi n + \phi_\text{CF})$, with $\phi_\text{CF}$ the emergent flux insertion experienced by CF) for different values of $L_y$ (Fig.~\ref{fig:central_charge}(c) illustrates $L_y=8$).

As shown in Fig.~\ref{fig:central_charge}(b), our results conform with an ordinary Fermi liquid (where $c = N_w$) iff the CF experience periodic (anti-periodic) boundary conditions at $L_y = 6$, $8$ ($L_y = 7$), respectively.

\begin{figure}[tp]
	\includegraphics[width=\columnwidth]{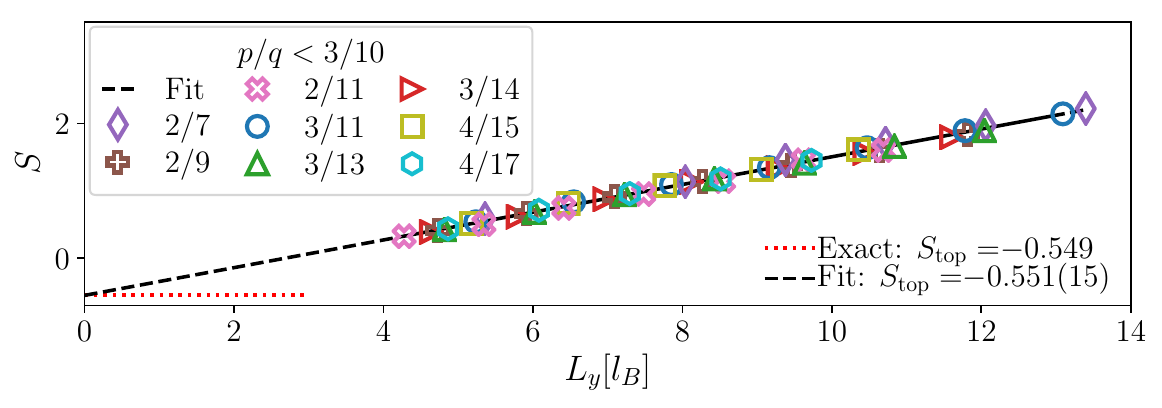} 
	\caption{Entanglement entropy in the Laughlin phase at $V=10$ as a function of cylinder circumference $L_y$ (in units of the magnetic length $\ell_B$) at $\chi=600$. 
	Results for cylinders of various $4 \leq L_y \leq 10$ and flux densities $n_{\phi} = p/q$ are combined. 
	\label{fig:S_topo}}
\end{figure}

We further substantiate the identification and stability of the Laughlin state, when the LLL-manifold has several magnetic sub-bands, i.e., $n_{\phi} = p/q$, $p>1$.
Firstly, we compute the entanglement entropy $S_E$ of bisected cylinders with varying $L_y$.
Unlike previous (i)DMRG simulations of the Harper-Hofstadter model, we simultaneously vary $n_{\phi}$, motivated by the computational advantages this brings in the torus geometry \cite{bauerQuantumGeometryStability2016, andrewsStabilityFractionalChern2018}.
This allows us to combine data for several $n_{\phi}$, providing a significant technical improvement upon earlier works. 
We obtain a much larger data set at reasonably low computational cost. 
Importantly, we find that the entanglement entropy depends on $L_y$ expressed in units of the magnetic length $\ell_B = \sqrt{\hbar/eB}=a \sqrt{q/2\pi p}$, so smaller values of $n_{\phi}$ allow for larger $L_y$ in units of lattice sites. 
Our data for $S_E(L_y/\ell_B)$ are shown in Fig.~\ref{fig:S_topo}, revealing a collapse onto a single line \footnote{We have excluded an outlier for the case of $p/q = 2/7, L_y = 5$, because its density profile differed significantly from that of a Laughlin state. 
We have also excluded results for the highest flux densities, $p/q \geq 3/10$ as experimental results suggest the Laughlin state is destabilized for $n_{\phi} \gtrsim 0.4$ \cite{hafeziFractionalQuantumHall2007}. }.
This data is consistent with the area law prediction of the entanglement entropy, $S_E = \alpha L_y + S_{\mathrm{topo}}$, with $S_{\mathrm{topo}}$ the topological entanglement entropy \cite{kitaevTopologicalEntanglementEntropy2006a}.
Extrapolation from our data yields $S_{\mathrm{topo}} = -0.551(15)$, accurately matching the prediction of $S_{\mathrm{topo}} = - \ln \sqrt{3} \approx -0.549$ for the $\nu = 1/3$ Laughlin state \cite{haqueEntanglementEntropyFermionic2007, lauchliEntanglementScalingFractional2010, zaletelTopologicalCharacterizationFractional2013a}. 
The collapsed data include cases where the LLL consists of $p=2,3,$ or $4$ bands, confirming that all of these cases consistently yield the topological nature of the Laughlin state, which is further supported by the respective entanglement spectra shown in Appendix C.

\emph{Discussion:} Given our data supporting a direct QH plateau transition between a CI state and FCI state in the Hofstadter model, we examine possible scenarios explaining the transition. 
We consider field theory descriptions for a continuous transition. 
Several works have considered CF flux-attachment transitions between different QH states  \cite{wenTransitionsQuantumHall1993,chenMottTransitionAnyon1993,yeCoulombInteractionsQuantum1998, groverQuantumPhaseTransition2013, barkeshliContinuousTransitionFractional2014, barkeshliContinuousPreparationFractional2015, leeEmergentMultiFlavorQED2018}. 
Transitions between a superconductor and a $\nu=1/2$ Laughlin state \cite{barkeshliContinuousTransitionFractional2014} and between different states in the same Jain series \cite{leeEmergentMultiFlavorQED2018, maEmergentMathrmQCDQuantum2020} seem most similar to our current setting. 
Both of these describe the transition as a Chern number changing transition experienced by a single fermionic parton band and the gap closing at multiple Dirac points simultaneously.

In our work, the CI phase does not require a parton construction, and is described by fermions in the lowest Hofstadter band with Chern number $C_1=4$. 
The FCI phase is described by two parton fields experiencing $C=1$ Chern bands to effectuate flux attachment, while the third parton field describes the effective physics of the CFs---also a $C=1$ Chern band for the Laughlin state. 
Hence, to describe the transition, one would have to explain the simultaneous change of the CF band Chern number, and fractionalisation of particles into three partons to give rise to the two additional fields. 
While effective theories in which this type of fractionalisation must occur at once may exist, a more generic case might display a sequence of several transitions affecting individual parton fields. 
Furthermore, our numerics indicate that the central charge at the critical point grows systematically with cylinder circumference (cf.~Fig.~\ref{fig:central_charge}).
While a theory of multiple Dirac points would predict central charges similar in magnitude to ours, it cannot explain the scaling of $c(L_y)$.
A single transition would thus require a more exotic explanation, such as the emergence of a 2D Fermi surface at the critical point \cite{gongChiralSpinLiquid2019}. 
The construction of theories for the interaction-driven CI-FCI plateau transition forms an exciting field of future study.

In conclusion, we have shown evidence of a direct phase transition between a $\nu = 1/3$ Laughlin state and an integer CI in the Harper-Hofstadter model at flux density $n_{\phi} = 3/11$.
Additionally, we used simultaneous scaling of cylinder circumference and flux density to obtain smooth data coverage of the cylinder measured in magnetic lengths and obtained an improved estimate for the topological entanglement entropy.

While we have provided a first step towards a theoretical explanation for the CI-FCI phase transition in terms of a 2D Fermi surface, it would be interesting to find a more complete theoretical description. 
Moreover, while the transition at $n_{\phi} = 3/11$ appears to be a direct phase transition, we have found some indications that at $n_{\phi} = 3/10$, the transition occurs via a yet uncharacterized intermediate phase. 
Future research could elucidate this latter transition by formulating a field theoretical model of flux attachment in the transition.

\begin{acknowledgments}
The authors would like to thank Nigel R. Cooper for helpful discussions, Steven H. Simon for comments on the manuscript, and Johannes Hauschild for support using TenPy \cite{hauschildEfficientNumericalSimulations2018}.
LS acknowledges funding from a University of Kent Vice Chancellor's Scholarship.
FP acknowledges the support of the DFG Research Unit FOR 1807 through grants no. PO 1370/2-1, TRR80, the Nanosystems Initiative Munich (NIM) by the German Excellence Initiative, the Deutsche Forschungsgemeinschaft (DFG, German Research Foundation) under Germany's Excellence Strategy -- EXC-2111-390814868 and the European Research Council (ERC) under the European Unions Horizon 2020 research and innovation program (grant agreement no.~771537).
GM acknowledges funding from the Royal Society under grants URF UF120157 and URF\textbackslash R\textbackslash 180004 and from the Engineering and Physical Sciences Research Council (EPSRC) under grant EP/P022995/1.
GM and FP acknowledge the hospitality of the Aspen Center for Physics, which is supported by National Science Foundation grant PHY-1607611. 
GM thanks the Theory of Condensed Matter Physics Group at the Cavendish Laboratory for their hospitality.
\end{acknowledgments}

\bibliography{multiband_hofstadter_letter.bib}

\begin{thebibliography}{88}
\bibitem [{spl()}]{splittingChernBands}%
  \BibitemOpen
  \href@noop {} {}\bibinfo {note} {Similarly, a parent Chern band could be
  split into multiple sub-bands by applying periodic superlattice potentials.
  Care would have to be taken to ensure that the resulting sub-bands do not
  overlap and gaps open fully.}\BibitemShut {Stop}%
\end{thebibliography}

\newpage
\appendix

\section{Appendix A: Details of the Laughlin state in multiple sub-bands}
We here present further details to demonstrate the existence and stability of the Laughlin state at flux densities $n_{\phi} = p/q$, $p>1$, i.e., where the lowest Landau level (LLL) consists of more than one sub-band. 
We include cases with $p=2$, $3$ and $4$ sub-bands in the LLL.
For each $p$, we consider several $q$.

\begin{figure}[t]
	\includegraphics[width=.8\columnwidth]{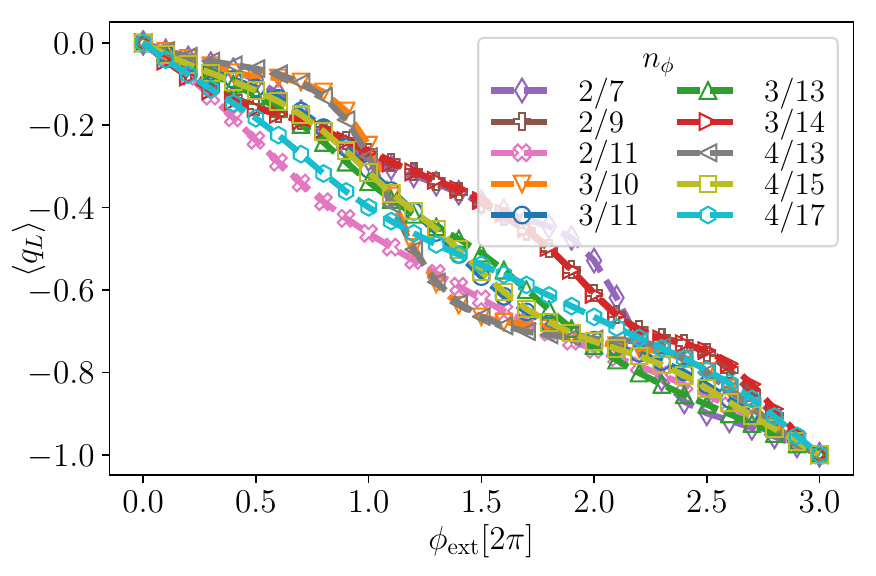} 
	\caption{Flux insertion for various flux densities $n_{\phi}$. In all cases, $L_y = 6$, $V=10$, $\chi=600$.}
	\label{fig:laughlin_pump}
\end{figure}

Firstly, Fig.~\ref{fig:laughlin_pump} shows the response to a flux pump procedure (as discussed in the main text) for all flux densities, on an $L_y = 6$ cylinder with $V=10$.
We observe that, although there is some variation in the precise course of the charge transport, in all cases a single charge quantum is transported upon adiabatic insertion of $3 \times 2 \pi$ flux.
Thus, all cases show a Hall response $\sigma_H = \Delta q / \Delta \phi_{\rm{ext}} = 1/3$, which is consistent with a $\nu = 1/3$ Laughlin state. 

\begin{figure}[t]
	\includegraphics[width=\columnwidth]{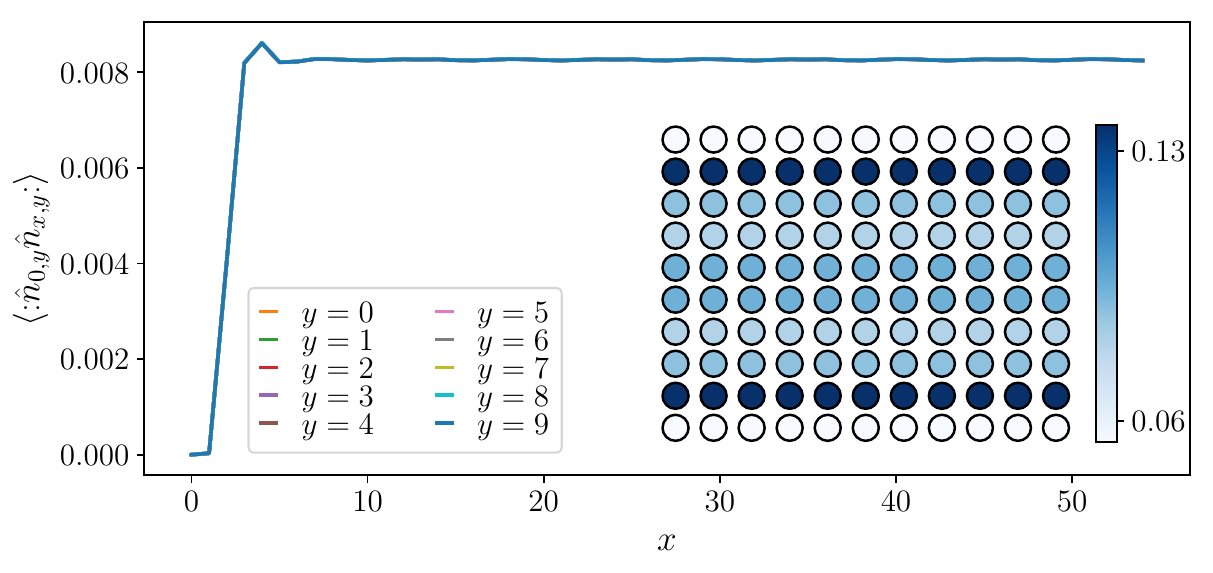} 
	\caption{Density-density correlation functions $\langle :\mathrel{\hat{n}_{0, y} \hat{n}_{x, y}}: \rangle$ along an $L_y = 10$ cylinder. The inset shows the density profile on a strip geometry (open boundary conditions in the $\hat{y}$-direction).}
	\label{fig:corr_density}
\end{figure}

Secondly, Fig.~\ref{fig:corr_density} shows both the density-density correlation functions as well as the density profile for $n_{\phi} = 3/11$ for $L_y = 10$ at $V = 10$.
The density-density correlations reveal a strong correlation hole at small $r$, followed by rapidly damped oscillations at intermediate $r$, before the asymptotic value representative of a homogeneous fluid is reached.
This is consistent with the Laughlin state, which is a liquid with no long range order, but with strong short-range correlations maximizing typical inter-particle distances.
With periodic boundary conditions along $\hat{y}$ (i.e., on the cylinder geometry), the density profile is homogeneous (not shown).
The inset in Fig.~\ref{fig:corr_density} therefore shows the density profile for a strip geometry (i.e., with infinite boundary conditions along $\hat{x}$, but open boundary conditions along $\hat{y}$) to demonstrate the effect of the edges of the system.
This density profile shows translational invariance along the cylinder, as well as oscillations around the expected bulk value and overshoot near the edge.
While some small finite size effects are visible in the bulk, all these results are consistent with a Laughlin state.

\begin{figure}[t]
	
	\includegraphics[width=\columnwidth]{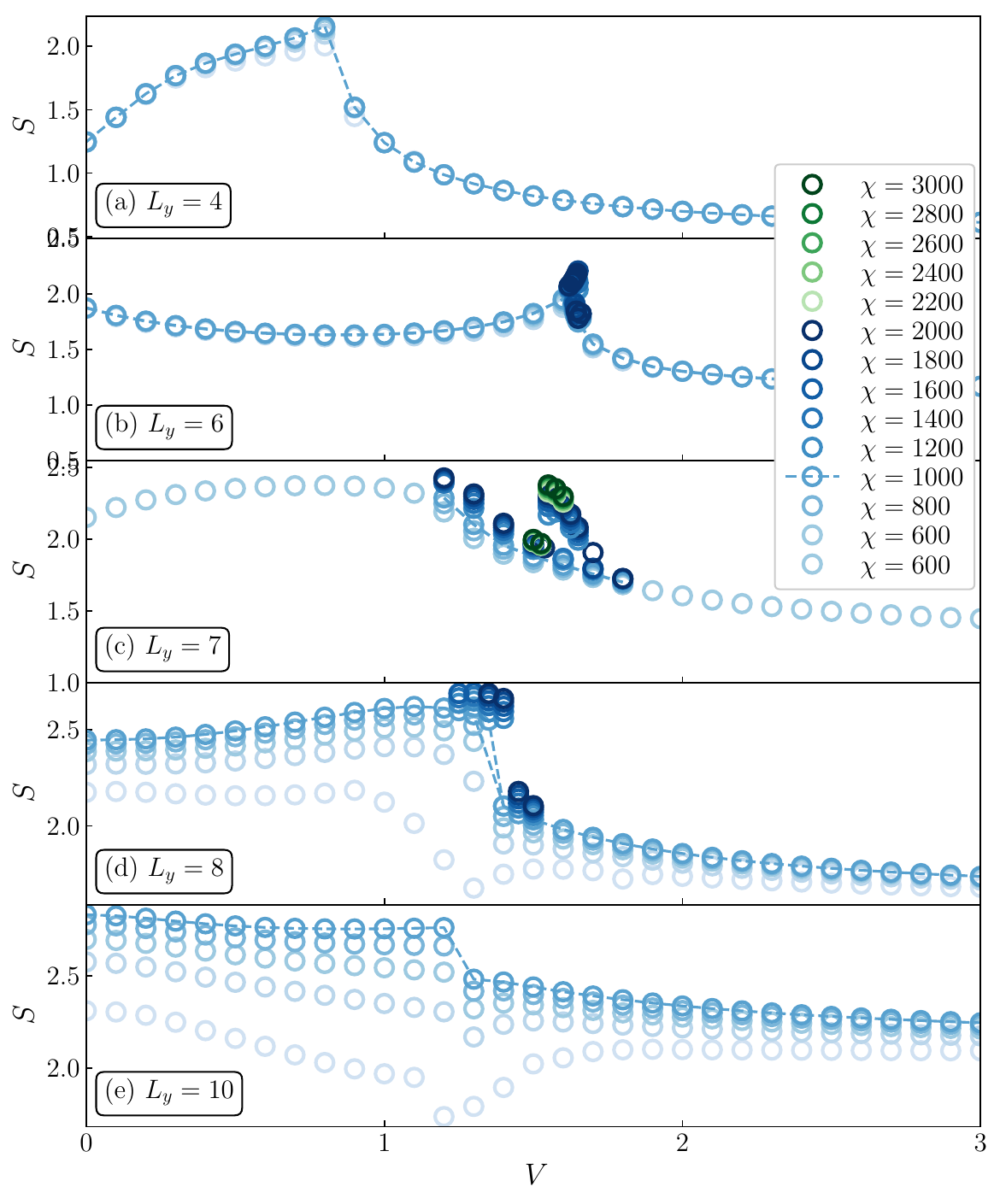} 
	\caption{Entanglement entropy (Von Neumann) $S$ as function of interaction strength $V$ and bond dimensions $\chi$ up to $\chi=1000$ throughout, and up to $\chi = 3000$ in the transition region, for (a) $L_y = 4$, (b) $L_y = 6$, (c) $L_y = 7$, (d) $L_y = 8$, (e) $L_y = 10$.}
	\label{fig:s_vs_v}
\end{figure}

Finally, Fig.~\ref{fig:s_vs_v} shows the scaling of the entanglement entropy with interaction $V$ and bond dimension $\chi$, for cylinders with $L_y = 4, 6, 7, 8$ and $10$.
The singular values $s_i$ for any bond in the Matrix Product State give the entanglement entropy as $S = - \sum_i s_i^2 \log s_i^2$, which we evaluate for a bond that corresponds to a bisection of the cylinder.
We note that with increasing $L_y$, it becomes increasingly difficult to resolve the physics at the transition, as evidenced by the strong growth of $S(\chi)$ with bond dimension.
Hence, we cannot fully converge the ground state wave function near the critical point, for $L_y\geq 6$.
In all cases, we observe signatures of a phase transition, where for bigger $L_y$ the observed features are less sharp than at smaller ones.
Near the transition, there is significant growth of $S$ as function of $\chi$, indicative of a critical point. This implies that DMRG convergence is challenging near the critical point and must be analyzed in terms of finite entanglement scaling. We include additional discussion of the high-bond dimension data for $L_y = 7$ and $L_y = 8$ in appendix D.

\section{Appendix B: Single-particle density matrix}
\begin{figure}[t]
	\includegraphics[width=\columnwidth]{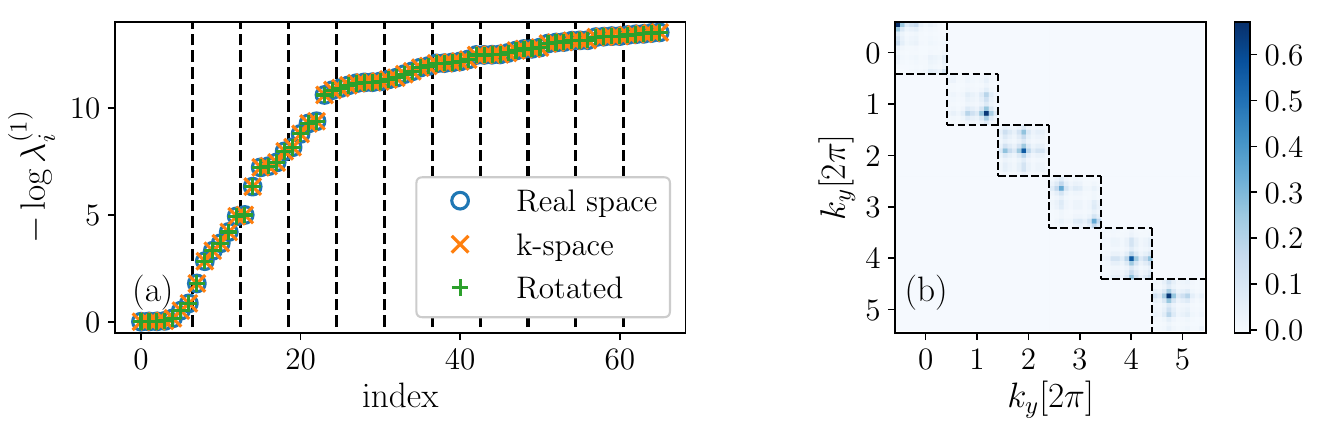} 
	\caption{The single-particle density matrix at $L_y = 6$.
		(a) The spectrum of the matrix remains unchanged under the transformations described in the main text, and equations \ref{eq:1pdm_fourier} and \ref{eq:1pdm_transformed}.
		Vertical dashed lines separate every $6$ consecutive values.
		(b) After the final reordering of the indices, the matrix is block-diagonal in $k_y$.
		Each block contains $11 \times 11$ values, due to the size (in $\hat{\vec{x}}$ of the MUC and MPS unit cell.
	\label{fig:density_matrix} }
\end{figure}

The single-particle density matrix $\rho_{ij} = \braket{\psi | c^{\dagger}_i c_j | \psi}$ can be transformed to momentum space using the Fourier transformation of the creation- and annihilation operators. 
Transforming only the $y$-coordinate, we have
\begin{equation} \label{eq:1pdm_fourier}
\hat{c}_{x, y} = \frac{1}{\sqrt{L_y}} \sum_{k_y = 0}^{L_y-1} e^{i 2 \pi k_y y / L_y} \hat{c}_{x, k_y},
\end{equation}
and thus
\begin{equation} \label{eq:1pdm_transformed}
\braket{c^{\dagger}_{x, k_y} c_{x' k_y'}} = \frac{1}{L_y} \sum_{y y'} e^{i 2 \pi (k_y y - k_y' y') / L_y} \braket{c^{\dagger}_{xy} c_{x'y'} }.
\end{equation}

We combine this basis transformation with a trivial reordering of the basis to make $k_y$ the dominant index, leading to a matrix that is block-diagonal in momentum sectors [Fig.~\ref{fig:density_matrix}(b)]. 
The spectrum of the matrix remains unchanged throughout this procedure [Fig.~\ref{fig:density_matrix}(a)].
This change of basis allows us to classify each eigenvalue by its $k_y$ sector, shown in terms of the coloring of symbols in Fig.~\ref{fig:transition_observables}(d) in the main text, and supporting our finding that the number of the most significant, low-lying density-matrix eigenvalues can be mapped to the eigenstates for the lowest $1/\nu$ single-particle bands.

\section{Appendix C: Momentum-resolved entanglement spectra}
\begin{figure}[tp]
	\includegraphics[width=\columnwidth]{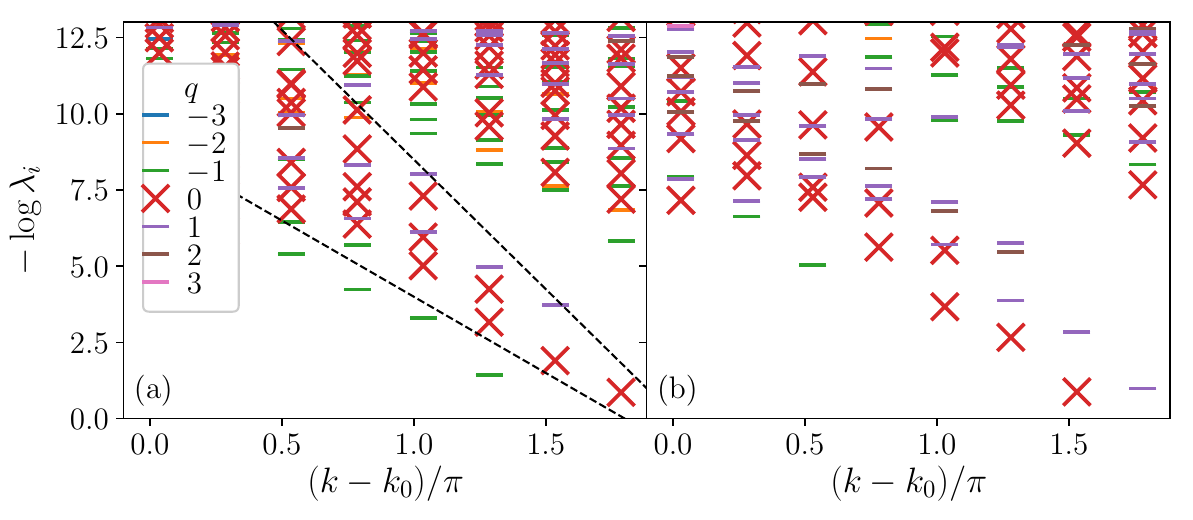} 
	\caption{Momentum-resolved entanglement spectrum for a bisected cylinder with $L_y = 8$, $\chi = 600$. 
		Colors denote different MPS quantum number sectors. 
		(a) $p/q = 1/3$, dashed lines indicate the semi-degeneracies of the first few levels within the $q=0$ sector. (b) $p/q = 3/11$. 
		Both spectra show the partition counting of semi-degenerate levels. \label{fig:kspecs}}
\end{figure}

To further probe the topological nature of the Laughlin state, we compute the momentum-resolved entanglement spectrum (ES) \cite{pollmannDetectionSymmetryprotectedTopological2012a, Cincio2013} for flux densities that we have studied, examples of which are shown in Fig.~\ref{fig:kspecs}. 
Momenta of the Schmidt eigenstates were extracted from the eigenvalues of translations around the cylinder $\hat T_y(a) \state{\chi_i^R} = e^{-i 2 \pi k_y / L_y} \state{\chi_i^R}$. The calculation proceeds by constructing a mixed transfer matrix of this operator inserted into the ground-state wave function to evaluate the matrix element $\bra{\chi_i^R}  \hat T_y(a) \state{\chi_i^R}$.

Our data clearly show the partition counting associated with the edge modes of the Laughlin state, expected to show up due to the bulk-boundary correspondence relating the ES to the edge modes' energy spectrum \cite{liEntanglementSpectrumGeneralization2008, qiGeneralRelationshipEntanglement2012}. 
In the lowest-lying charge sector, we find a clear signature of the universal $U(1)$ edge counting of the number of states in consecutive momentum sectors given by the partition counting sequence $1$, $1$, $2$, $3$, $5$, $7,\ldots$. 
For the charge-neutral sector, our data show reveal the four-five momentum sectors satisfying this counting of low-lying quasi-degenerate states, before these are no longer clearly distinguishable from the continuum above the entanglement gap. 
In the adjacent charged charge sectors, the entanglement energies tend to be higher overall, so it becomes more difficult to discern the edge from bulk states.

\section{Appendix D: Quantum Hall plateau transition at $L_y = 7$ and $L_y = 8$}
\label{app:largerLy}
\begin{figure}[tp]
	\includegraphics[width=\columnwidth]{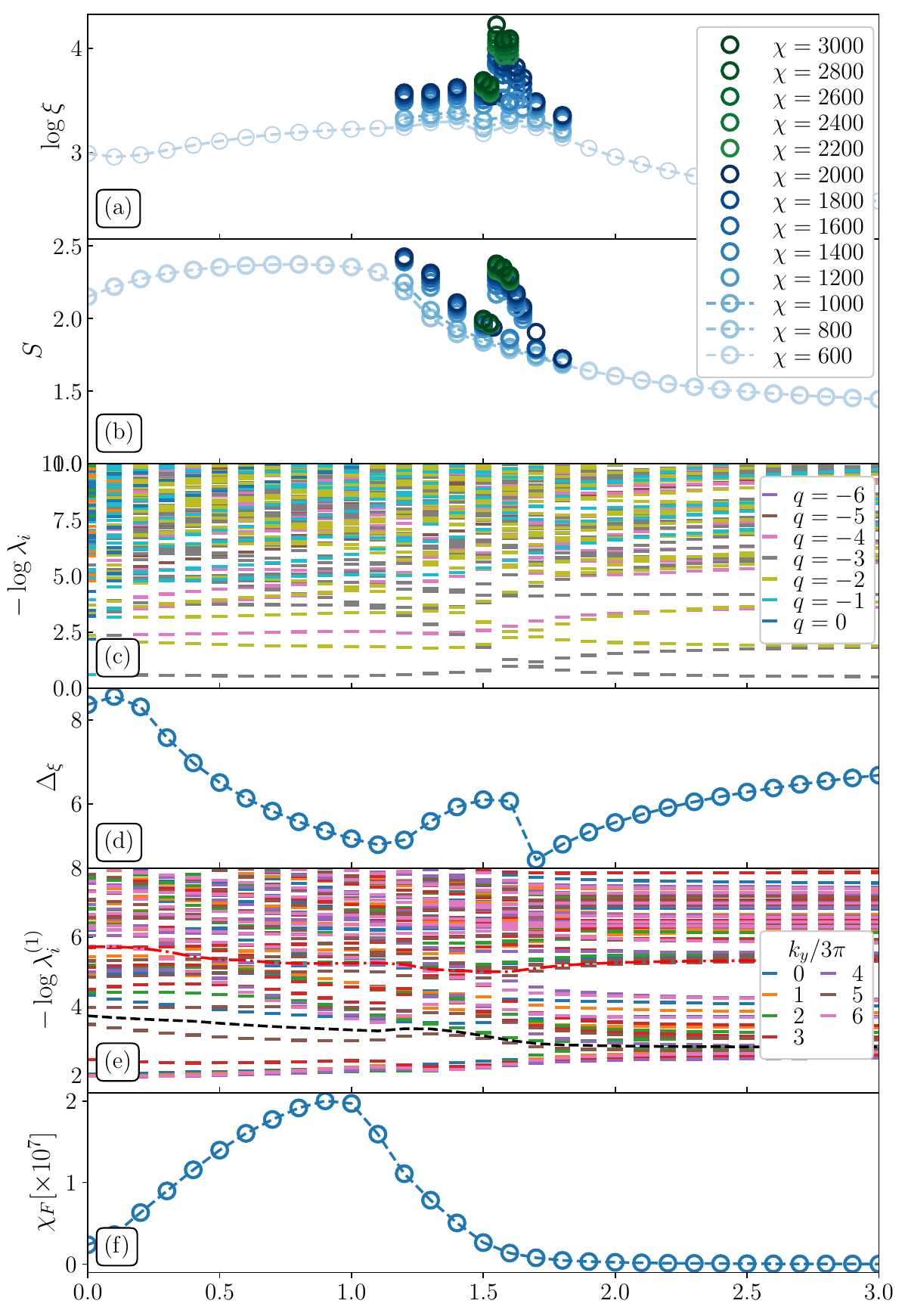} 
	\caption{The plateau transition at $L_y = 7$.
		Shown are:		
		(a)~Correlation length in lattice sites (the circumference $\log L_y \approx 1.95$ falls below the range shown.) and (b)~entanglement entropy for several values of the bond dimension $600 \leq \chi \leq 1000$ throughout, and increasing to $\chi = 2000$ near the transition and up to $\chi=3000$ between $V=1.5$ and $V=1.6$.
		(c)~Momentum-space ES (colours denote different quantum number sectors). 
		(d)~Entanglement gap. 
		(e)~Spectrum of the single-particle density matrix. The dashed black and dash-dotted red lines denote the lowest 7 and 21 levels, respectively.
		(f)~Fidelity susceptibility $\chi_F$.
		Subfigures (c) through (f) have $\chi=600$.
		}
	\label{fig:transition_Ly7}
\end{figure}

\begin{figure}[tp]
	\includegraphics[width=\columnwidth]{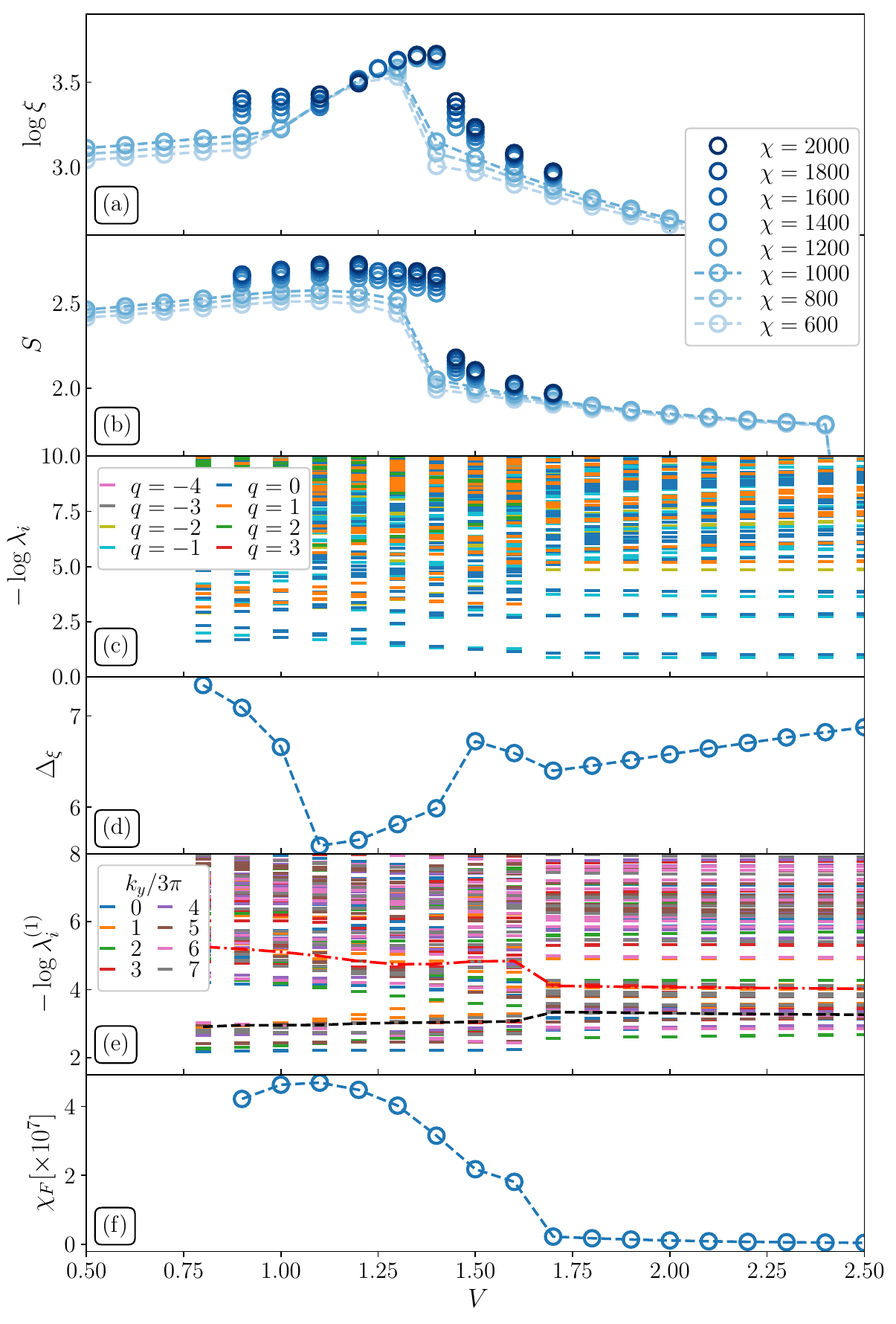} 
	\caption{The plateau transition at $L_y = 8$.
		Shown are:		
		(a)~Correlation length in lattice sites (the circumference $\log L_y \approx 2.089$ falls below the range shown) and (b)~entanglement entropy for several values of the bond dimension $600 \leq \chi \leq 1000$ throughout and up to $\chi = 2000$ near the transition.
		(c)~Momentum-space ES (colours denote different quantum number sectors). 
		(d)~Entanglement gap. 
		(e)~Spectrum of the single-particle density matrix. The dashed black and dash-dotted red lines indicate the lowest $8$ and $24$ levels, respectively.
		(f)~Fidelity susceptibility $\chi_F$.
		Subfigures (c) through (f) have $\chi=600$.
	}
	\label{fig:transition_Ly8}
\end{figure}

In the main text, we discuss the plateau transition between a $C = +4$ Chern insulator and a $\nu = 1/3$ Laughlin state at $n_\phi = 3/11$, at a circumference $L_y = 6$.
At that circumference, the transition appears most clearly.
Here, we show additional results for $L_y = 7$ (Fig.~\ref{fig:transition_Ly7}) and $L_y = 8$ (Fig.~\ref{fig:transition_Ly8}).
The flux insertion procedure has not been repeated for these circumferences due to the large computational cost associated with the required pseudo-adiabatic evolution of the wave function.
All other observables shown in the main text are included here.

\begin{figure}[tp]
	\includegraphics[width=\columnwidth]{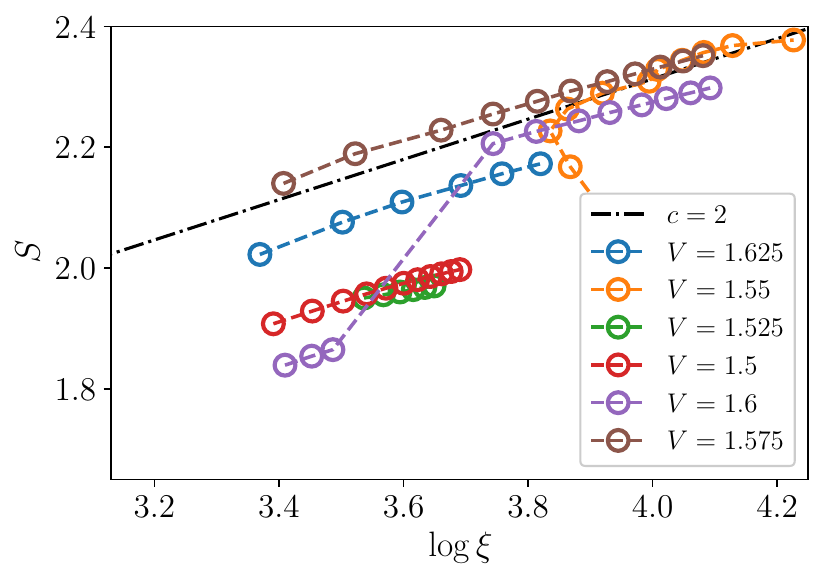} 
	\caption{Finite entanglement scaling for the plateau transition at $L_y = 7$, demonstrating the emergence of a linear relation between the entanglement entropy $S$ and the logarithm of the correlation length $\xi$ is consistent with a central charge of $c=2$ near the critical point of $V_c \simeq 1.55$. The underlying data include simulations up to bond dimension $\chi=3000$ at each interaction strength.
	}
	\label{fig:scaling_Ly7}
\end{figure}

We observe that examined with an MPS of the same bond dimension, the transition appears less sharp at $L_y = 7$ and even less so for $L_y=8$ than it does at $L_y = 6$. Overall, the transition region is broader than for $L_y=6$ in both these geometries. However, upon increasing the bond dimension, we see sharper features appear in the entanglement and correlation length at $L_y=7$. 

We first broadly identify the transition region by probing for the largest continued increase of the correlation length and entanglement entropy with MPS bond dimension.
For $L_y = 7$ [Fig.~\ref{fig:transition_Ly7}(a-b)], we see that both $\xi(\chi)$ and $S(\chi)$ go through a relatively broad plateau for interactions up to $V=1.2$ and $V=1.5$, respectively, before falling off towards the FCI phase. 
However, on top of this plateau, we find an additional feature linked to a steep rise in both quantities for $\chi>1400$ and located between $V=1.5$ and $V=1.6$. 
In order to further investigate this feature, we have run calculations up to a bond dimension of $\chi=2000$ for the region surrounding the transition, and up to $\chi=3000$ between $V=1.5$ and $V=1.6$. We have analysed these high-$\chi$ data to verify the simultaneous scaling of $\log (\chi)$ and $S_E(\chi)$ with the bond dimension, and find a good fit of linear scaling in the throughout the transition region. A detail of these data are is shown in Fig.~\ref{fig:scaling_Ly7}. We see that linear relation corresponding to $S_E = c/6 \log(\xi)$ provides a good description of the data: The maximum entropy and correlation lengths are seen at $V=1.55$, our data point closest to criticality. DMRG convergence has been noticeably slowed down at this point compared to nearby interaction strengths, and some numerical noise is still visible. In particular, this DMRG run crossed over from a lower-entropy state to a higher entropy one during subsequent increases of the bond dimension. Between $\chi=2200$ and $\chi=2800$ the data are fit with a slope of $\frac{1}{6}\times 1.98$, while the $\chi=3000$ datapoint lies slightly below this line.
At the slightly lower interaction of $V=1.5$, a linear fit of $S_E$ vs $\log(\xi)$ matches the central charge $c=2$ when restricted to bond dimensions below $\chi=2000$ (data shown in the main text), but then tails off slightly at larger bond dimension. Interestingly, our calculation at the interaction strength $V=1.525$ has instead converged to a state that yields a scaling consistent with $c=1$. According to our analysis in terms of a CF Fermi surface state, both $c=1$ and $c=2$ could describe the critical theory, corresponding to two possible values of $0$ or $\pi$ for the emergent flux felt by composite fermions in the ground state. This emergent flux could be selected randomly depending on the random initial state used in our DMRG runs, so both outcomes are consistent with our proposed scenario for a critical point. We also note that calculations at lower bond dimensions shown in Fig.~\ref{fig:transition_Ly7}(b) tend to show a dip in the correlation length for $V\simeq V_c$, which could be related to the alternative lower central charge $c=1$ branch. By contrast, our higher bond dimension calculations overwhelmingly are consistent with $c=2$ scaling (cf.~Fig.~\ref{fig:scaling_Ly7}), so we believe this reflects the dominant state for the critical behaviour. At larger interaction strength of $V=1.6$, we find the slope is noticeably smaller, and is best fit by a non-integer value of $c\simeq 1.54$, scaling linearly up to the highest bond dimensions we explored. The non-integer value reflects that this interaction strength is situated outside the critical region. Overall, these findings match the phenomenology for a critical point with central charge $c=2$, as we would expect to see continued linear scaling with the bond dimension only precisely at the critical point. With increasing distance from the critical point, we instead expect the growth in entanglement to tail off once the bond dimension is sufficient to accurately capture the finite correlation length of the off-critical system, as seen in these data.
Altogether, these features allow us to situate a transition described by a $c=2$ central charge in the close vicinity of $V=1.55$.

For $L_y = 8$ [Fig.~\ref{fig:transition_Ly8}(a-b)], we find that the correlation length goes through a broad maximum which is reached at $V=1.3$ to $V=1.4$, depending on the bond dimension $\chi$. 
At larger $V$, we then find a rapid drop of the correlation length, and to the right of the maximum, there is a very strong dependence of the correlation length on $\chi$. 
Furthermore, the precise location of this drop in $\xi$, which one may take as an estimate of the numerically determined transition point $V_c(\chi)$, moves towards larger values as the bond dimension is increased. 
We believe that this shift is due to a competition between the two states for entanglement resources and that it can be explained by noting that the higher-entanglement Chern insulator phase becomes more favourable compared to the fractional Chern phase once the MPS can hold sufficient entanglement to represent it sufficiently accurately. 
In other words, the DMRG algorithm has a bias towards the FCI phase at low bond dimension, given that the CI state has a higher intrinsic entanglement entropy than the FCI state. 
At larger $V$, the correlation length continues to evolve smoothly towards lower values. 
Unlike the data for $L_y=6$, or $7$, we did not identify any sharp peaks in $\xi$, but our numerical results are not sufficiently closely spaced on the $V$-axis to exclude this possibility, and even larger bond dimensions would likely be required to reveal any detailed structure. 
The entanglement entropy in [Fig.~\ref{fig:transition_Ly8}(b)] shows very similar behaviour to the correlation length, but the peak is overall slightly broader and values of $S$ start sloping down before undergoing a jump to lower values associated with the FCI phase between $V=1.4$ and $V=1.45$.

The broad behaviour for $L_y=7$, $8$ is similar to what we observed at $L_y=6$: Within the transition region, both the correlation length $\xi(\chi)$ and the entanglement entropy $S(\chi)$ [Fig.~\ref{fig:transition_Ly7}(a,b), Fig.~\ref{fig:transition_Ly8}(a,b)] show a continual growth with bond dimension, which is consistent with a divergence in the bond dimension, as well as with $\abs{V-V_c}$. 
The region near the transition shows several marked features, with a sharp peak emerging at $L_y=7$ and the most prominent feature at $L_y=8$ being a sudden drop in the two quantities towards the right side of the transition. 
We believe that both are consistent with a continuous quantum phase transition occurring at $V_c$. 
The scaling of the entanglement entropy with correlation length, shown in Fig.~\ref{fig:central_charge} and Fig.~\ref{fig:scaling_Ly7}, further corroborates this view. 
The data shown in Fig.~\ref{fig:central_charge} were taken at the interaction strengths of $V_\text{scale}^{L_y=6}=1.634$, $V_\text{scale,1}^{L_y=7}=1.5$ (orange squares) and $V_\text{scale}^{L_y=8}=1.45$, respectively.
These points are located within the transition region and slightly to the FCI side of an estimated critical interaction strength, where we see clear linear scaling up to $\chi=2000$. Additionally, we showed the scaling for $L_y=7$ and $V_\text{scale,2}^{L_y=7}=1.55$ (orange stars) up to bond dimension $\chi=3000$. 

The additional data in Fig.~\ref{fig:transition_Ly7}(c-f) and Fig.~\ref{fig:transition_Ly8}(c-f) show a number of entanglement spectral properties and the fidelity susceptibility of the ground state wave function for a lower bond dimension of $\chi=600$. 
Here, the underlying calculations were performed while semi-adiabatically increasing the interaction strength $V$ in order to obtain a consistent allocation of charge quantum numbers between neighbouring interaction strengths (running the DMRG from random initial states can yield different members of the degenerate set of ground states, which have different allocations of charge sectors).
We have confirmed that these results are not dependent on the direction of change in $V$, i.e., that no hysteresis occurs.

Where the momentum-resolved entanglement spectrum exhibited a discontinuous change at $L_y = 6$, for both $L_y = 7$ and $L_y = 8$ we observe a more continuous evolution of the spectrum in the transition region, which also includes some level crossings.
For example, in Fig.~\ref{fig:transition_Ly7}(c) we see the lowest two entanglement eigenstates cross near $V=1.6$, and we further clarify that the respective states are located in different momentum sectors (while they carry the same charge quantum number), so the crossing is protected by translation symmetry. 
In Fig.~\ref{fig:transition_Ly8}(c,e) we note that the entanglement spectrum and single-particle density matrix eigenvalues display a significant change between $V=1.6$ and $V=1.7$, and then remain nearly unchanged for larger values of $V$. Based on this information, we would locate the transition at a slighly larger value of $V_c$ than what we have determined from the entanglement entropy and correlation length.

Sharp discontinuities in the entanglement gap [Fig.~\ref{fig:transition_Ly7}(d) and Fig.~\ref{fig:transition_Ly8}(d)] can be ascribed to the gap suddenly being measured from a different ground state (and thus in a different momentum- and quantum number sector), rather than a physical effect. In particular, the level crossing discussed in the previous paragraph explains the sharp feature seen at the corresponding interaction strength in the entanglement gap in Fig.~\ref{fig:transition_Ly7}(d).
The spectra of the single-particle density matrix in [Fig.~\ref{fig:transition_Ly7}(e) and Fig.~\ref{fig:transition_Ly8}(e)] show a first gap at a number of eigenstates matching the number of states in the lowest band to the left of the transition, while a second gap emerges adjacent to thrice that number of states to the right of the transition. At $L_y=7$ the first gap closes smoothly with $V$ and the second one opens gradually above the transition. Al $L_y=8$, a number of states seem to descend into the first gap on approaching the transition from low $V$, but the second gap remains at a fairly constant value above the transition. Neither oft these cases show any other prominent gaps, again pointing towards a direct transition from the CI to FCI states. We should caution that these data were taken at bond dimension of $\chi=600$, and corrections to the higher-lying states are expected at larger bond dimension in the transition region.
Results for the fidelity susceptibility $\chi_F$ [Fig.~\ref{fig:transition_Ly7}(f) and Fig.~\ref{fig:transition_Ly8}(f)] again do not demonstrate sharp features, but do suggest a broad region where the ground state varies rapidly with the interaction $V$.

\end{document}